\documentclass[a4paper,11pt]{article}
\usepackage{pos}
\usepackage{enumitem,multirow}

\title{Exploring the high-energy gamma-ray spectra of TeV blazars}
 \ShortTitle{High-energy gamma-ray spectra of TeV blazars}

\author*[a]{Qi Feng}

\affiliation[a]{Physics Department of Physics and Astronomy, Barnard College, Columbia University, \\ 
New York, NY 10027, USA}

\forColl{VERITAS} 

\emailAdd{qifeng@nevis.columbia.edu}

\abstract{The highest-energy blazars exhibit non-thermal radiation extending beyond 1 TeV with high luminosities and strong variabilities, indicating extreme particle acceleration in their relativistic jets. The gamma-ray spectra of blazars contain information about the distribution and cooling processes of high-energy particles in jets, the extragalactic background light between the source and the observer, and potentially, the environment of the gamma-ray emitting region and exotic physics that may modify the opacity of the universe to gamma rays. We use data from \textit{Fermi}-LAT and VERITAS to study the variability and spectra of a sample of TeV blazars across a wide range of gamma-ray energies, taking advantage of more than ten years of data from both instruments. The variability in both the GeV and TeV gamma-ray bands is investigated using a Bayesian blocks method to identify periods with a steady flux, during which the average gamma-ray spectra, after correcting for the pair absorption effect from propagation, can be parameterized without the risk of mixing different flux states. We report on the search for intrinsic spectral curvature and spectral variability in these blazars, in an effort to understand the physical mechanisms behind the high-energy gamma-ray spectra of TeV blazars.}

\FullConference{37$^{\rm{th}}$ International Cosmic Ray Conference (ICRC 2021)\\
		July 12th -- 23rd, 2021\\
		Online -- Berlin, Germany}

\begin{document}
\maketitle

\section{Introduction}
Blazars are active galactic nuclei (AGN) with relativistic jets pointing along the line of sight to the observer. The broadband spectral energy distribution (SED) of blazars exhibits a two-bump structure with a first component peaking at infrared to keV energies and a second one at MeV to TeV energies. 
The radiative processes in blazar jets involve relativistic particles (leptonic and possibly hadronic), although many questions remain unresolved~\cite{blandford2018}, such as the gamma-ray emission mechanism. 
The small viewing angle of the jet makes it possible to observe strong relativistic effects, such as a boosting of the emitted power and a shortening of the characteristic time scales. 

The energy spectra of blazars are crucial in determining the physical processes underlying jet acceleration and radiation. Considering the largest section of the GeV to TeV blazar peak possible increases the ability to distinguish between competing theoretical predictions. Access to spectra over the full MeV to TeV range makes it possible to address, among others, the following questions:
\noindent
\begin{itemize}
\item{} What are the highest particle energies that are realized in relativistic jets? 
\item{} Is there internal absorption of $\gamma$-rays in the emission region? 
\item{} Can the MeV-TeV spectra be consistently described as a single emission component, or is there evidence for multiple spectral components? 
\item{} Are the different flux states in blazars due to continuous variation of the same underlying physical phenomenon, or do bright flares show evidence of a new emission region and particle population?
\end{itemize}

The study of blazars has benefited in the past decade from the availability of \textit{Fermi}-LAT observations in the energy band from 50 MeV - 1 TeV and imaging atmospheric Cherenkov telescope (IACT) array observations in the energy band above $\sim$100 GeV, smoothly covering the GeV-TeV energy band. However, joint studies between the two types of instruments are vulnerable to biases, particularly when differing source variability in the two energy bands is not taken into account. Such a bias can be addressed by selecting time intervals in which the source flux does not vary significantly in either band. 

The SEDs of high-frequency-peaked BL Lac objects (HBLs) peak at energies where \textit{Fermi}-LAT and the IACTs are sensitive, making them good candidates for the studying the questions posed above via joint spectral analysis. While spectral analysis in the \textit{Fermi}-LAT range alone has provided insight into these questions, extension to higher energies promises more power to test theoretical predictions~\cite{romoli2016}.


\section{VERITAS and \textit{Fermi}-LAT observations}
This ongoing study uses a sample of 17 strongly detected high-frequency-peaked BL Lac objects (HBLs). In this work, we show results on four of the 17 sources: 1ES 1218+304, 1ES 1011+496, 1ES 1959+650, and 1ES 2344+514. All sources have well-established redshifts, allowing for confident estimation of their spectra after correction for absorption by the extragalactic background light~\cite[EBL;][]{ebl}. In addition to having measured redshifts, the sources were selected for their brightness in both the \textit{Fermi}-LAT and VERITAS energy bands, with the goal of minimizing the statistical uncertainties in the joint spectral analysis. The sources have been observed by both VERITAS and \textit{Fermi}-LAT for over a decade, facilitating a long baseline variability analysis. As all sources are observed to be variable by both VERITAS and \textit{Fermi}-LAT, variability analysis is a critical component of this study.

The VERITAS and \textit{Fermi}-LAT instrument performance is described in detail elsewhere (see \cite{VERITAS} and \cite{Fermi}). Data were processed with the standard calibration and reconstruction pipelines for each instrument. Throughput corrections were applied to the VERITAS data to address instrument aging~\cite{throughputcorrections}. The key properties of the sources and the strength of their detection by VERITAS and \textit{Fermi}-LAT are given in Table~\ref{sourceproperties}.

\begin{table}
\centerline{
\begin{tabular}{ccccccc}
Target & Redshift & Observing time window & $\sigma_{\textrm{VTS}}$  & $\sigma_{\textrm{LAT}}$ & $N_{blocks}(VTS)$ & $N_{blocks}(LAT)$ \\
\hline
\hline
1ES 2344+514  & 0.044 & 2007.10 - 2015.12 & 29.1 & 62.7  & 9 & 1 \\
1ES 1959+650 & 0.048 & 2007.11 - 2016.06 & 70.1  & 163 & 13 & 10 \\
1ES 1218+304 & 0.182 & 2008.12 - 2018.06 & 64.9  & 77.4 & 10 & 6 \\
1ES 1011+496 & 0.212  & 2008.01 - 2018.02 & 40.7 & 166 & 7 & 7 \\
 \hline
\end{tabular}
}
\caption{\small{Sources analyzed for this study, together with redshift, observing time, detection significance, and the number of Bayesian blocks by VERITAS and \textit{Fermi}-LAT.}}
\label{sourceproperties}
\end{table}

\section{Spectral analysis}

\begin{figure}
{\includegraphics[width=0.49\textwidth]{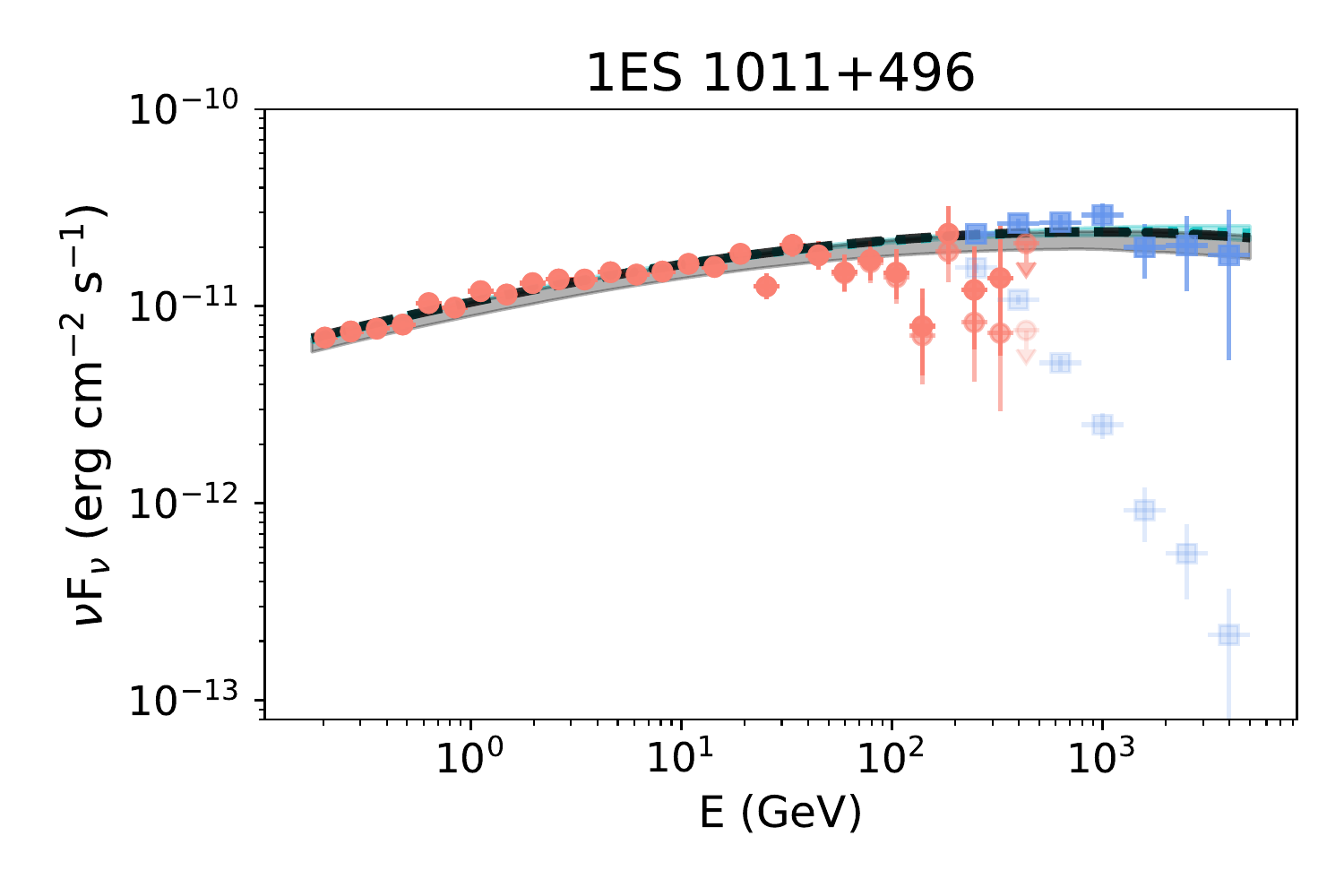}}
{\includegraphics[width=0.49\textwidth]{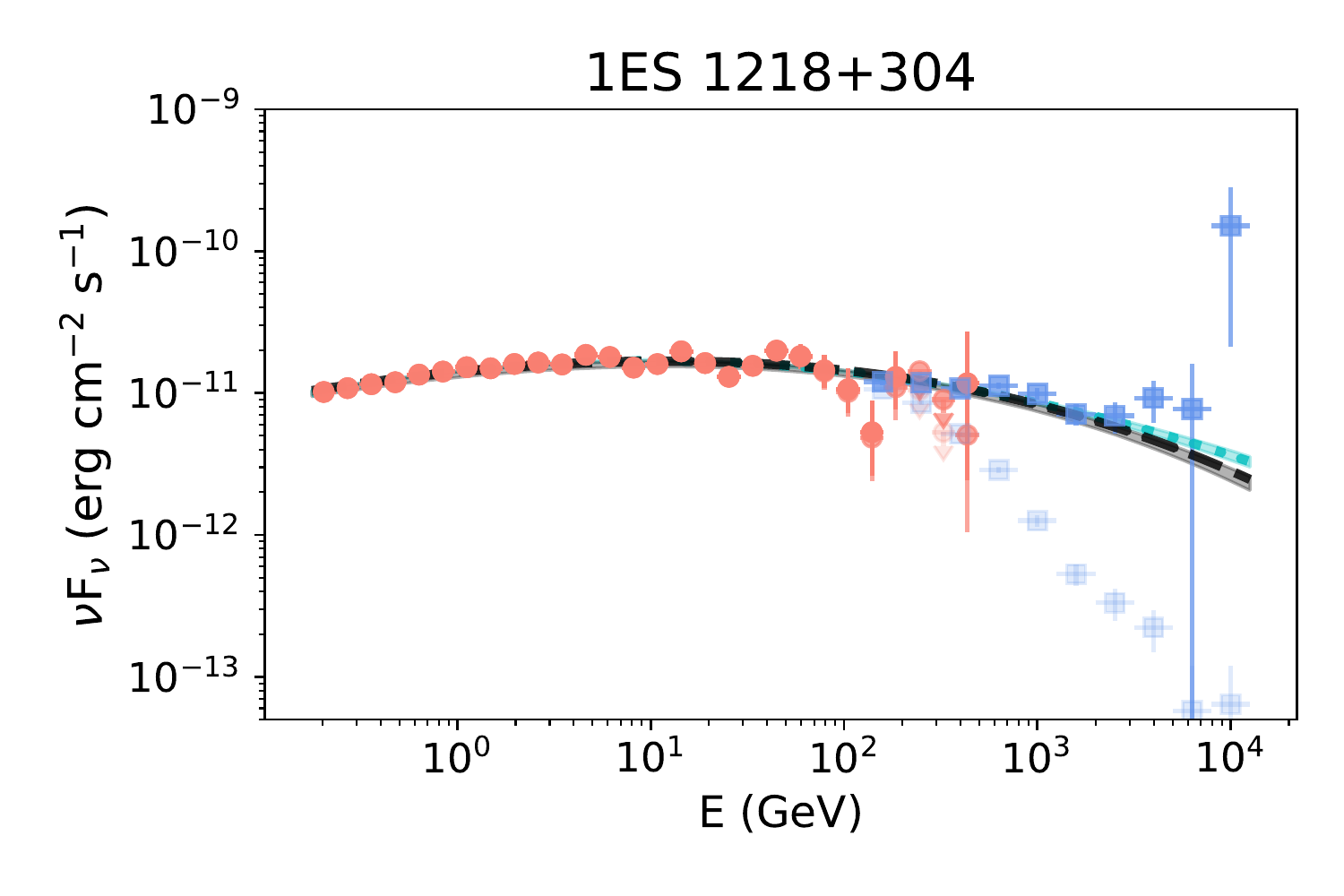}} \\
{\includegraphics[width=0.49\textwidth]{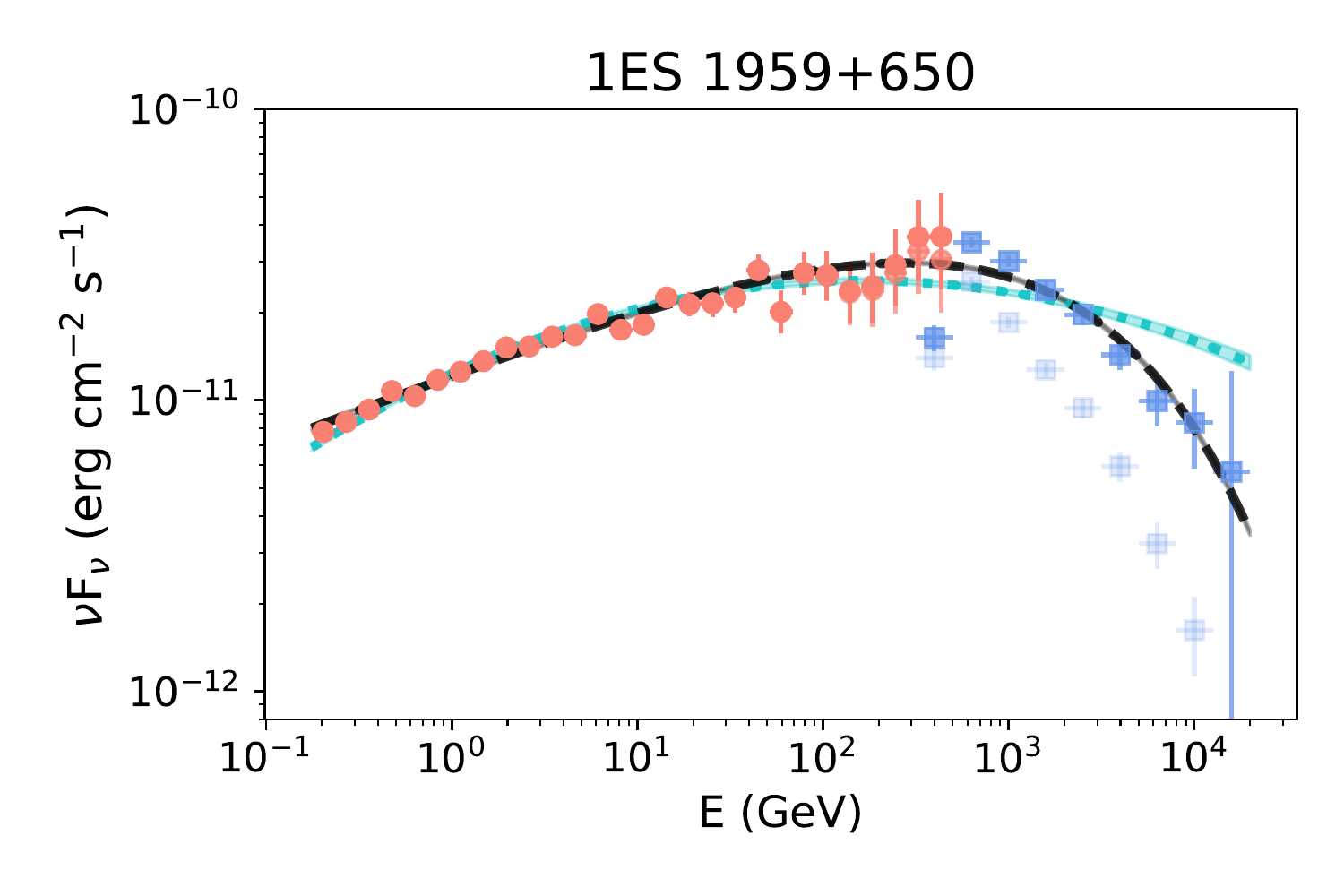}}
{\includegraphics[width=0.49\textwidth]{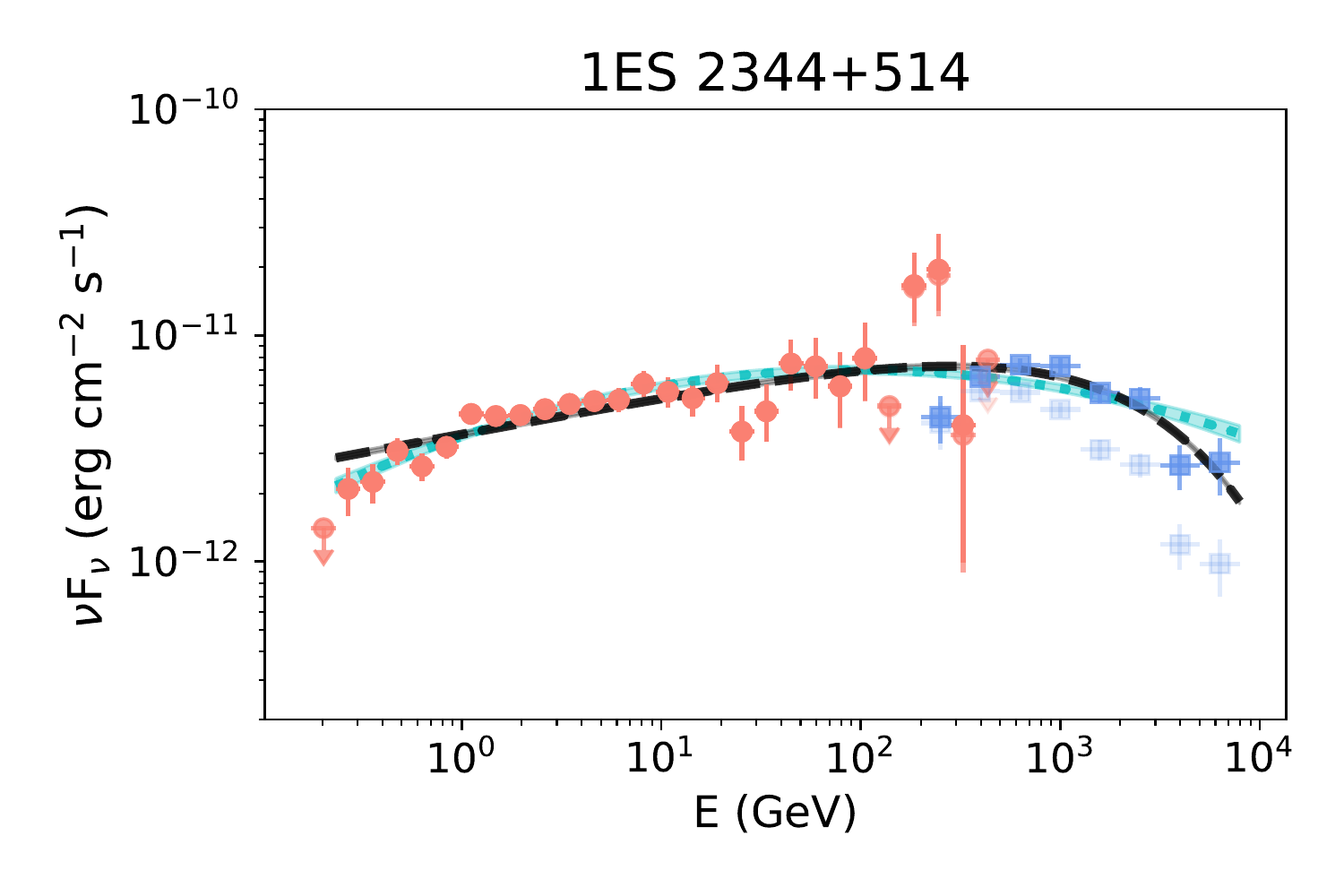}}

\caption{\small{Time-averaged \textit{Fermi}-LAT (red filled circles) and VERITAS (blue filled squares) spectra of the four blazars. From top left to bottom right, the spectra shown are for 1ES 1011+496, 1ES 1218+304, 1ES 1959+650, and 1ES 2344+514. The faded colors represent the observed spectra, whereas the deeper colors represent the deabsorbed spectra using the EBL model from \cite{ebl}. The best-fit power law with an exponential cutoff is shown for each blazar in black dashed lines, and the log-parabola models are shown in cyan dotted lines. }}\label{fig:specs}
\end{figure}

The spectra averaged over the entire \textit{Fermi}-LAT and VERITAS observing time windows are constructed and are shown in Figure~\ref{fig:specs}. With the longest possible exposure, these spectra provide the strongest statistical constraints on the gamma-ray emission, although the interpretations are limited due to the lack of consideration of variability. 
To study the intrinsic spectral curvature of the sources (or potential exotic effects from the propagation of gamma rays), we corrected for the effect of EBL before fitting the spectra. 

All of the four sources show hard spectra in the energy range of \textit{Fermi}-LAT and spectral curvature after correcting for the absorption from EBL. 
Two models were used to describe the spectra, namely a power-law model with an exponential cutoff 
\begin{equation} \label{eq1}
\frac{dN}{dE} = N_0(E/E_0)^{-\Gamma} \exp{[-(E/E_\text{c})^{\beta_\gamma}}],
\end{equation}
where $N_0$ is the normalization at the energy $E_0$, $\Gamma$ is the photon index, $E_\text{c}$ is the cutoff energy, and $\beta_\gamma$ quantifies how steep the cutoff is;
and a log-parabola model
\begin{equation} \label{eq2}
\frac{dN}{dE} = N_0(E/300~\text{GeV})^{[-\alpha-\beta \log_{10}(E/300~\text{GeV})]},
\end{equation}
where $\alpha$ is the index and the $\beta$ is the curvature parameter. 
The exponential cutoff shape in the first model is related to the particle distribution and emission mechanism \cite{romoli2016}, while the log-parabola model could suggest stochastic acceleration \cite{Berg2019}. 

When the power-law model with an exponential cutoff 
is used to describe the spectra, all four sources show a sub-exponential, or stretched, cutoff ($\beta_\gamma$<1). 1ES 1959+650 exhibits the most prominent spectral curvature among the four sources, which is better described by a power law with a sub-exponential cutoff than a log-parabola model. %
For the other three sources, there are no significant differences between the two models. 
The gamma-ray spectral cutoff ($\beta_\gamma$) can be used to estimate the spectral cutoff ($\beta_e$) of the energy distribution of the emitting particles, which sheds light on the acceleration and energy loss rate of these particles. For example, for a source emitting through the synchrotron self-Compton (SSC) mechanism in the Thomson regime, the cutoff sharpness $\beta_e$ of the radiating electrons can be calculated as $\beta_e = 4\beta_\gamma/(1-\beta_\gamma)$ \cite{romoli2016}. For 1ES~1959+650, a $\beta_\gamma \sim 0.44$ gives a sharp electron cutoff $\beta_e \sim 3.1$ under the SSC scenario. 

\section{Variability analysis}

\begin{figure}
\centerline{\includegraphics[width=0.75\textwidth]{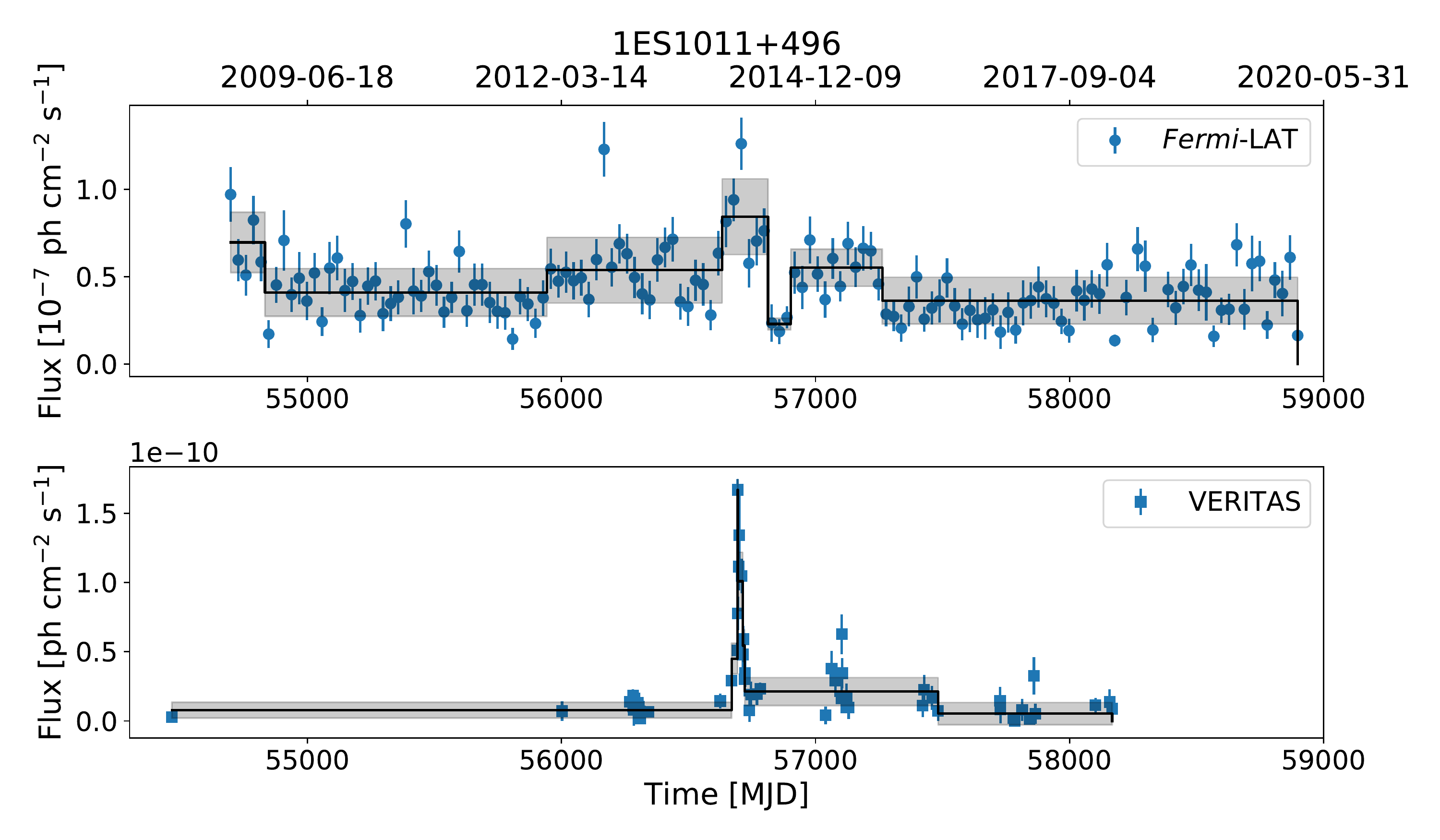}}
\centerline{\includegraphics[width=0.75\textwidth]{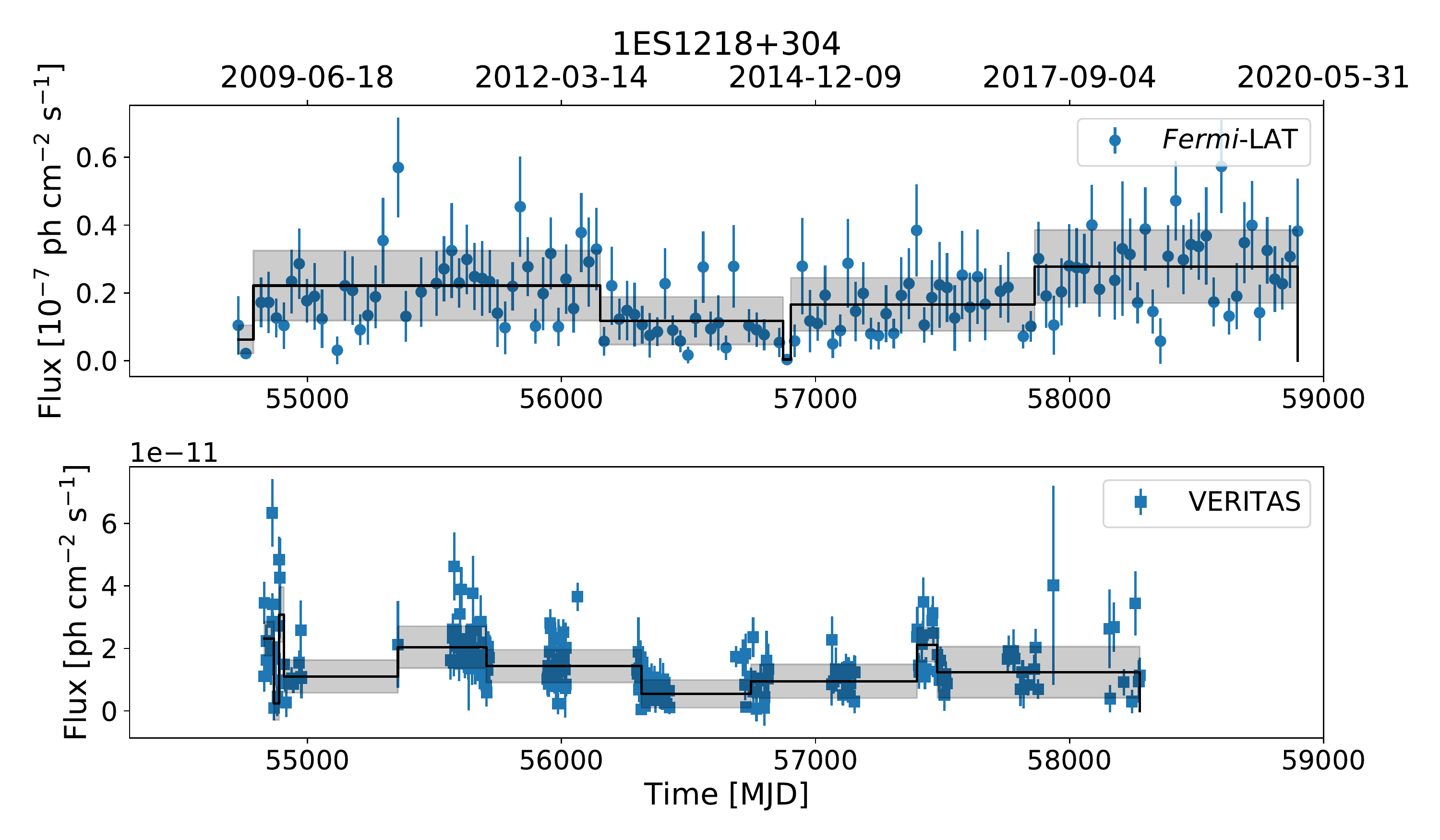}}
\centerline{\includegraphics[width=0.75\textwidth]{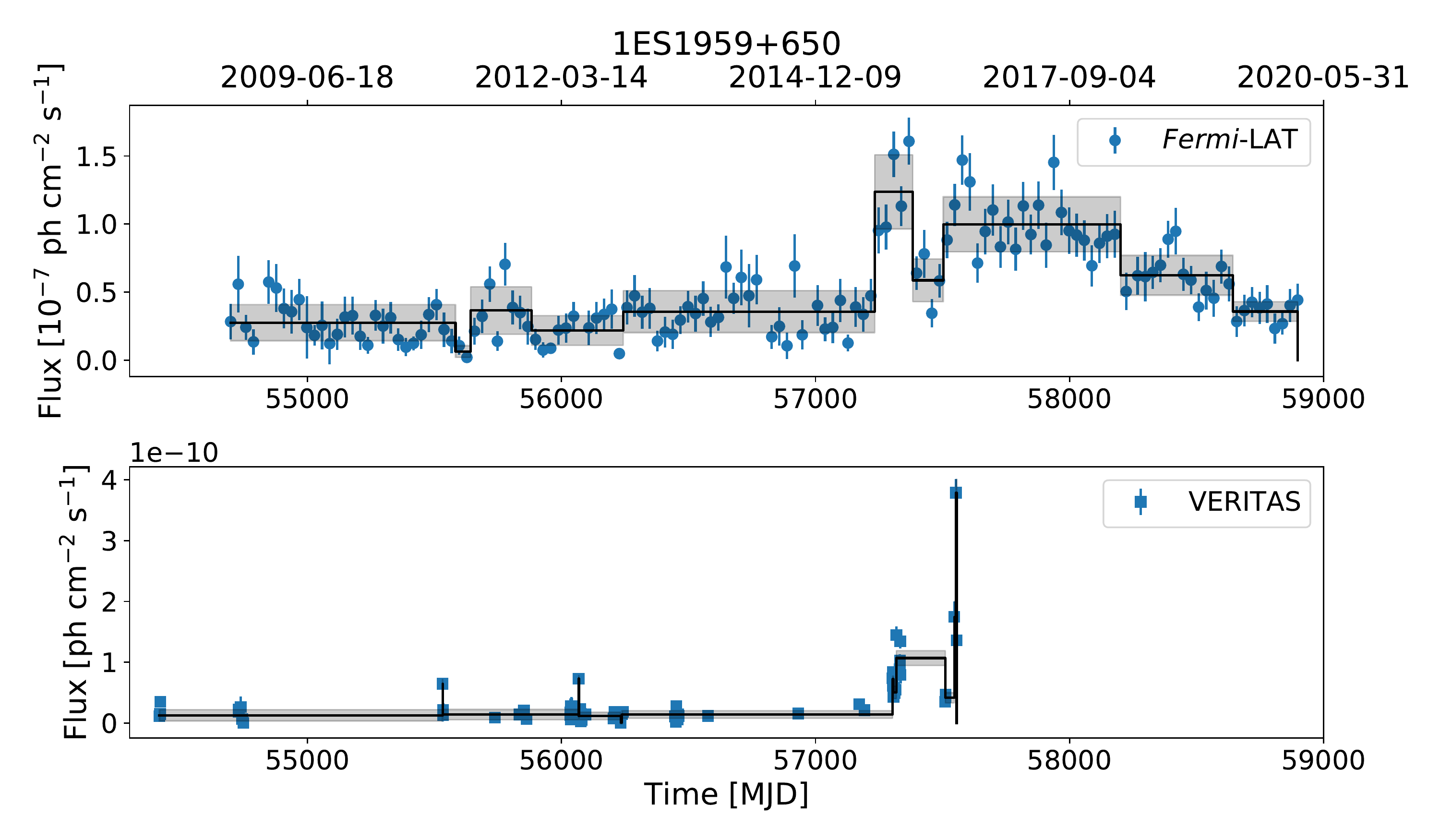}}
\caption{\small{The monthly \textit{Fermi}-LAT (top panels) and nightly VERITAS (bottom panels) light curves for three of the four sources. From top to bottom, the sources are 1ES 1011+496, 1ES 1218+304, and 1ES 1959+650, respectively. The black solid lines and the grey shaded regions show the average and the standard deviation of the fluxes within a Bayesian block. }}\label{fig:lightcurves}
\end{figure}

Light curves for three blazars were constructed for both the VERITAS and \textit{Fermi}-LAT observations and are shown in Figure~\ref{fig:lightcurves}. \textit{Fermi}-LAT lightcurves are shown for an energy threshold of 100 MeV in monthly time bins. VERITAS lightcurves were constructed in 24 hour time bins above an energy threshold of 200 GeV for all sources other than 1ES 1959+650. The high declination of 1ES 1959+650 with respect to the VERITAS site results in a higher energy threshold of 316 GeV.  

A Bayesian block analysis was performed on all lightcurves in order to define time intervals in which the flux did not vary significantly~\cite{bayesblocks}. The significance of change points defining the boundary between two blocks of differing flux was required to be greater than 3$\sigma$. The blocks are shown in Figure~\ref{fig:lightcurves}, and the number of Bayesian blocks per source is shown in Table~\ref{sourceproperties}. All three sources shown have significant variability in both energy bands. For 1ES 2344+514 (not shown), variability was significantly detected only in the TeV band and not in the GeV band due to its low GeV flux. 

In addition to defining periods of non-varying flux with the VERITAS and \textit{Fermi}-LAT energy bands, time intervals where neither VERITAS nor \textit{Fermi}-LAT observed flux variability were defined. These intervals of common non-varying flux are intended to prevent different flux states from being averaged together when performing joint spectral analysis, which can lead to a more stretched spectral cutoff. 

\section{Spectral variability analysis}
\begin{figure}
\centerline{\includegraphics[width=1.\textwidth]{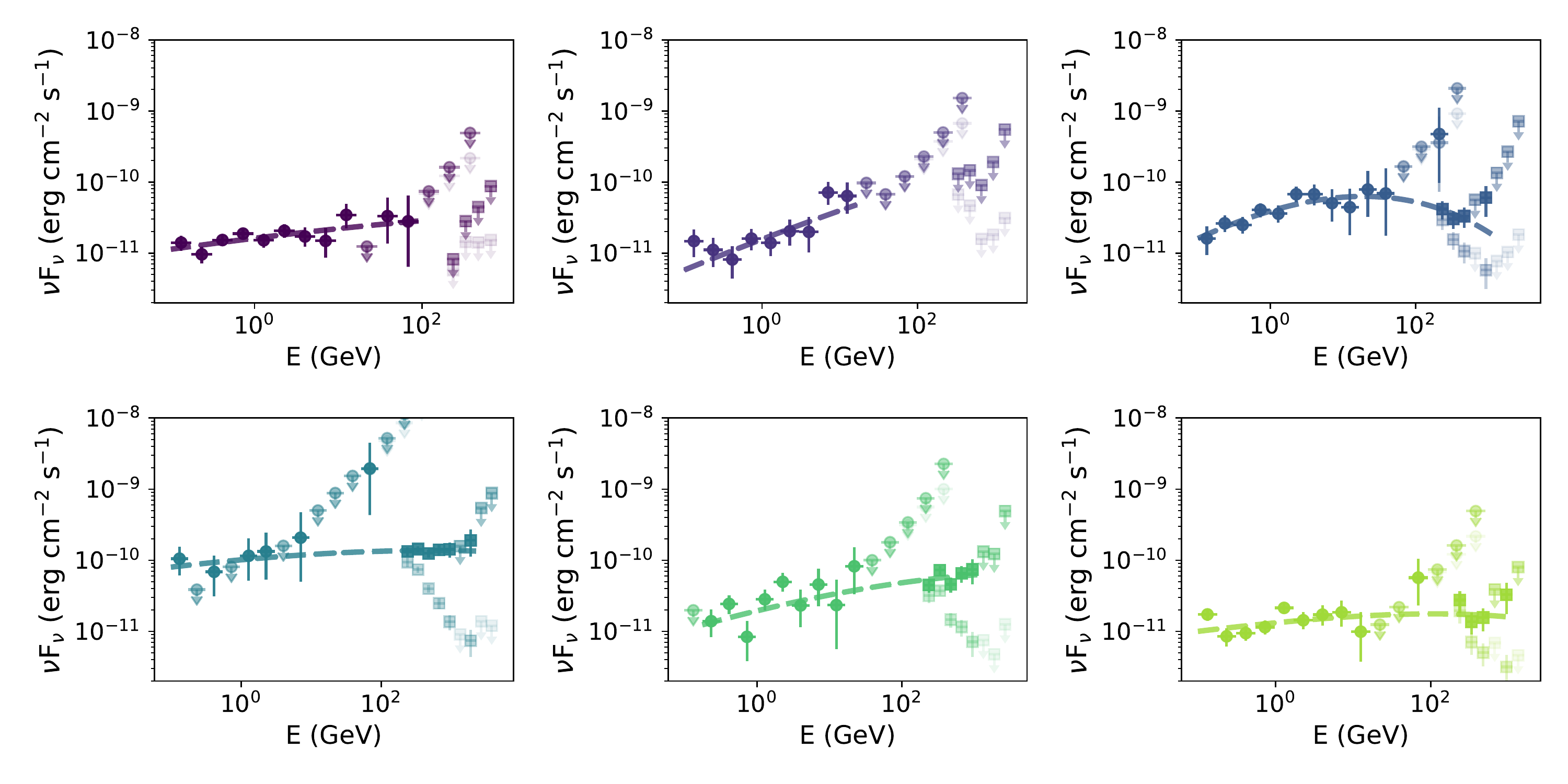}}
\caption{\small{Spectrum of 1ES 1011+496 integrated over six intervals determined by the Bayesian blocks around the time of the TeV gamma-ray flare. From top left to bottom right, the integration intervals progress in time. Filled circles and squares represent \textit{Fermi}-LAT and VERITAS spectra, respectively. The faded colors represent the observed spectra, whereas the deeper colors represent the deabsorbed spectra using the EBL model from \cite{ebl}. The dashed lines show the best-fit power-law model with an exponential cutoff without consideration of the upper limits. }}\label{fig:bbspec}
\end{figure}

The source 1ES 1011+496 exhibited a strong TeV flare in 2014, with contemporary elevated GeV flux. 
The spectrum of 1ES 1011+496 integrated over six intervals determined by both the VERITAS and \textit{Fermi}-LAT Bayesian blocks around the time of the TeV gamma-ray flare are shown in Figure~\ref{fig:bbspec}. 
When the flux of the source was the highest (lower left panel), the duration of the interval was also the shortest ($\sim1$ day). The large effective area of the VERITAS array allows good statistics in the TeV gamma-ray regime over such a short flaring episode, whereas the GeV gamma-ray spectrum was not as constraining. The opposite occurs for the longer intervals, during which the flux of the source was lower but the GeV spectrum is well constrained thanks to the long exposure from \textit{Fermi}-LAT enabled by its large field-of-view. 

The same analysis will be performed for the 17 sources in the sample. The relation between the spectral hardness and the flux of the sources will also be studied. 

\section{Summary}

In this work we present preliminary results from a study of gamma-ray spectral data of four TeV blazars, measured by Fermi-LAT and VERITAS, covering a decade in time. Blazars detected at TeV energies are perhaps the most extreme objects in the AGN population, the properties of which are not completely understood. The spectral data cover more than five decades in energy and time-averaged spectra show sub-exponential spectral curvature for all four sources. Significant variability was detected in the light curves. Analysis is ongoing to cover all hard-spectrum HBLs with a firm detection by VERITAS, and spectra for these sources will be assembled for the time windows during which both GeV and TeV fluxes are steady. These time periods will be selected following the same Bayesian block variability analysis as presented in this work. The block-by-block spectra will allow the particle distribution for a given steady state to be probed without mixing states, which introduces bias to the cutoff sharpness. We hope that the assembled GeV-TeV spectral data will help quantify the spectral hardness and curvature 
from $\sim 100$ MeV to $>10$ TeV and will provide insights into the acceleration and cooling mechanisms of the emitting particles in the jets of blazars.

\section*{Acknowledgements}
This research is supported by grants from the U.S. Department of Energy Office of Science, the U.S. National Science Foundation and the Smithsonian Institution, by NSERC in Canada, and by the Helmholtz Association in Germany. This research used resources provided by the Open Science Grid, which is supported by the National Science Foundation and the U.S. Department of Energy's Office of Science, and resources of the National Energy Research Scientific Computing Center (NERSC), a U.S. Department of Energy Office of Science User Facility operated under Contract No. DE-AC02-05CH11231. We acknowledge the excellent work of the technical support staff at the Fred Lawrence Whipple Observatory and at the collaborating institutions in the construction and operation of the instrument.

%
%
%

\clearpage \section*{Full Authors List: \Coll\ Collaboration}

\scriptsize
\noindent
C.~B.~Adams$^{1}$,
A.~Archer$^{2}$,
W.~Benbow$^{3}$,
A.~Brill$^{1}$,
J.~H.~Buckley$^{4}$,
M.~Capasso$^{5}$,
J.~L.~Christiansen$^{6}$,
A.~J.~Chromey$^{7}$, 
M.~Errando$^{4}$,
A.~Falcone$^{8}$,
K.~A.~Farrell$^{9}$,
Q.~Feng$^{5}$,
G.~M.~Foote$^{10}$,
L.~Fortson$^{11}$,
A.~Furniss$^{12}$,
A.~Gent$^{13}$,
G.~H.~Gillanders$^{14}$,
C.~Giuri$^{15}$,
O.~Gueta$^{15}$,
D.~Hanna$^{16}$,
O.~Hervet$^{17}$,
J.~Holder$^{10}$,
B.~Hona$^{18}$,
T.~B.~Humensky$^{1}$,
W.~Jin$^{19}$,
P.~Kaaret$^{20}$,
M.~Kertzman$^{2}$,
T.~K.~Kleiner$^{15}$,
S.~Kumar$^{16}$,
M.~J.~Lang$^{14}$,
M.~Lundy$^{16}$,
G.~Maier$^{15}$,
C.~E~McGrath$^{9}$,
P.~Moriarty$^{14}$,
R.~Mukherjee$^{5}$,
D.~Nieto$^{21}$,
M.~Nievas-Rosillo$^{15}$,
S.~O'Brien$^{16}$,
R.~A.~Ong$^{22}$,
A.~N.~Otte$^{13}$,
S.~R. Patel$^{15}$,
S.~Patel$^{20}$,
K.~Pfrang$^{15}$,
M.~Pohl$^{23,15}$,
R.~R.~Prado$^{15}$,
E.~Pueschel$^{15}$,
J.~Quinn$^{9}$,
K.~Ragan$^{16}$,
P.~T.~Reynolds$^{24}$,
D.~Ribeiro$^{1}$,
E.~Roache$^{3}$,
J.~L.~Ryan$^{22}$,
I.~Sadeh$^{15}$,
M.~Santander$^{19}$,
G.~H.~Sembroski$^{25}$,
R.~Shang$^{22}$,
D.~Tak$^{15}$,
V.~V.~Vassiliev$^{22}$,
A.~Weinstein$^{7}$,
D.~A.~Williams$^{17}$,
and 
T.~J.~Williamson$^{10}$\\
\noindent
$^1${Physics Department, Columbia University, New York, NY 10027, USA}
$^{2}${Department of Physics and Astronomy, DePauw University, Greencastle, IN 46135-0037, USA}
$^3${Center for Astrophysics $|$ Harvard \& Smithsonian, Cambridge, MA 02138, USA}
$^4${Department of Physics, Washington University, St. Louis, MO 63130, USA}
$^5${Department of Physics and Astronomy, Barnard College, Columbia University, NY 10027, USA}
$^6${Physics Department, California Polytechnic State University, San Luis Obispo, CA 94307, USA} 
$^7${Department of Physics and Astronomy, Iowa State University, Ames, IA 50011, USA}
$^8${Department of Astronomy and Astrophysics, 525 Davey Lab, Pennsylvania State University, University Park, PA 16802, USA}
$^9${School of Physics, University College Dublin, Belfield, Dublin 4, Ireland}
$^{10}${Department of Physics and Astronomy and the Bartol Research Institute, University of Delaware, Newark, DE 19716, USA}
$^{11}${School of Physics and Astronomy, University of Minnesota, Minneapolis, MN 55455, USA}
$^{12}${Department of Physics, California State University - East Bay, Hayward, CA 94542, USA}
$^{13}${School of Physics and Center for Relativistic Astrophysics, Georgia Institute of Technology, 837 State Street NW, Atlanta, GA 30332-0430}
$^{14}${School of Physics, National University of Ireland Galway, University Road, Galway, Ireland}
$^{15}${DESY, Platanenallee 6, 15738 Zeuthen, Germany}
$^{16}${Physics Department, McGill University, Montreal, QC H3A 2T8, Canada}
$^{17}${Santa Cruz Institute for Particle Physics and Department of Physics, University of California, Santa Cruz, CA 95064, USA}
$^{18}${Department of Physics and Astronomy, University of Utah, Salt Lake City, UT 84112, USA}
$^{19}${Department of Physics and Astronomy, University of Alabama, Tuscaloosa, AL 35487, USA}
$^{20}${Department of Physics and Astronomy, University of Iowa, Van Allen Hall, Iowa City, IA 52242, USA}
$^{21}${Institute of Particle and Cosmos Physics, Universidad Complutense de Madrid, 28040 Madrid, Spain}
$^{22}${Department of Physics and Astronomy, University of California, Los Angeles, CA 90095, USA}
$^{23}${Institute of Physics and Astronomy, University of Potsdam, 14476 Potsdam-Golm, Germany}
$^{24}${Department of Physical Sciences, Munster Technological University, Bishopstown, Cork, T12 P928, Ireland}
$^{25}${Department of Physics and Astronomy, Purdue University, West Lafayette, IN 47907, USA}

\end{document}